\documentclass[preprint,10pt,a4paper,times,onecolumn]{elsarticle}
\pdfoutput=1
\usepackage{amssymb}
\usepackage{a4wide}

\newcounter{bla}

\usepackage{listings}
\usepackage{color}
\usepackage{textcomp}
\definecolor{listinggray}{gray}{0.9}
\definecolor{lbcolor}{rgb}{0.9,0.9,0.9}
\usepackage[bookmarks,colorlinks=false,%
pdftitle={ISICSoo: a class for the calculation of ionization cross sections from ECPSSR and PWBA theory},%
pdfauthor={Matej Batic},%
pdfborder={0 0 0},%
pdfstartview={Fit}]{hyperref}
\usepackage[english]{babel}
\usepackage{dcolumn}
\usepackage{ctable}
\usepackage{multirow}

\journal{Computer Physics Communications}
\begin{document}

\begin{frontmatter}
    
\title{ISICSoo: a class for the calculation of ionization cross sections from ECPSSR and PWBA theory} 
    
\author[infn,ijs]{Matej Bati\v{c}\corref{author}}
\author[infn]{Maria Grazia Pia}
\author[creighton]{Sam J.\ Cipolla}
    
\cortext[author] {Corresponding author.\\\textit{E-mail address:}
  matej.batic@ge.infn.it} 
\address[infn]{Instituto Nazionale di Fisica Nucleare, Sezione di
  Genova, Via Dodecaneso 33, 16146 Genova, Italy}  
\address[ijs]{Jo\v{z}ef Stefan Institute, Jamova 39, 1000 Ljubljana,
  Slovenia} 
\address[creighton]{Physics Department, Creighton
  University, Omaha, NE 68178, USA}

\begin{abstract}
ISICS, originally a C language program for calculating K-, L- and
M-shell ionization and X-ray production cross sections from ECPSSR and
PWBA theory, has been reengineered into a C++ language class, named
ISICSoo. The new software design enables the use of ISICS functionality
in other software systems. The code, originally developed for Microsoft
Windows operating systems, has been ported to Linux and Mac OS platforms
to facilitate its use in a wider scientific environment. The
reengineered software also includes some fixes to the original
implementation, which ensure more robust computational results and a
review of some physics parameters used in the computation. The paper
describes the software design and the modifications to the
implementation with respect to the previous version; it also documents
the test process and provides some indications about the software
performance.
\end{abstract}


\end{frontmatter}
  
  

{\bf NEW VERSION PROGRAM SUMMARY}

\begin{small}
\noindent
%
%
%
{\em Program Title:} ISICSoo \\
{\em Journal Reference:}  \\
{\em Catalogue identifier:} \\
{\em Licensing provisions:} Standard CPC licence, \url{http://cpc.cs.qub.ac.uk/licence/licence.html} \\
{\em Programming language:} C++ \\
{\em Computer:} 80486 or higher-level PC or Mac \\
{\em Operating system:} any OS with gcc compiler version 4.1 (or
newer); tested on Scientific Linux 5 (gcc 4.1.2), Mac OS X 10.6.5
(gcc  4.2.1) and Windows XP (MS Visual C++ 2010 Express) \\  
{\em Classification:}  16.7 \\
{\em Catalogue identifier of previous version:} ADDS\_v4\_0 \\
{\em Journal reference of previous version:} Comput.\ Phys.\ Comm.\
180 (2009) 1716 \\
{\em Does the new version supersede the previous version?:} no \\ 
{\em Nature of problem:} Ionization and X-ray production cross section 
calculations for ion-atom collisions.\\   
{\em Solution method:} Numerical integration of form factor using a
logarithmic transform and Gaussian quadrature, plus exact integration
limits. \\ 
{\em Reasons for the new version:} %
Capability of using ISICS physics functionality in other software
systems; porting the software to other platforms than Microsoft Windows;
improved computational robustness and performance. \\ 
{\em Summary of revisions:} %
Reengineering into a C++ class; several internal modifications to improve
correctness and robustness; updated binding energies tabulations;
performance improvements. \\
{\em Running time:} The running time depends on the selected atomic shell
and the number of polynomials used in the Gaussian quadrature 
integration. \\

\end{small}

\section{Introduction}
\label{intro}

ISICS (Inner-Shell Ionization Cross Sections)
\cite{isics,isics2006,isics2008,isics2011} is a C language program that
computes ionization and X-ray production cross-sections for K-, L-, and
M-shells in the ECPSSR \cite{ecpssr} and PWBA \cite{pwba} theoretical
frameworks, using exact integration limits for calculating the form
factors. Both theories have found use in numerous applications, such as
those involving elemental composition analysis or ion beam transport,
and research involving projectile charge change and energy loss, recoil
ion production, or target vacancy rearrangement and X-ray production. 

A few other open source programs calculating ECPSSR cross sections can
be found in the literature. \v{S}mit's code \cite{smit} calculates cross
sections for K- and L-shells and is implemented in Pascal. The ERCS08
program \cite{horvat}, implemented in FORTRAN, calculates cross section
for K-, L- and M-shell ionization according to various options of the
ECPSSR theory. Both of them, as well as ISICS, are stand-alone
applications; as such, they are not meant to be used within another
application. All these codes run on Windows platforms; ERCS08 can run
with minor modifications on other platforms equipped with a FORTRAN
compiler, but its graphical user interface (GUI) is specific to the
Windows environment. 

ISICSoo addresses these shortcomings: its main motivation is the
capability of using ISICS physics functionality in other software
systems, rather than only as a stand-alone program, and in a variety of
platforms, not limited to the original Microsoft Windows
environment. For this purpose, ISICS has been reengineered into a C++
class, which preserves the same physics functionality as the original
stand-alone ISICS program, while providing greater flexibility of use
and improved robustness of implementation.   

Although the physics capabilities of ISICSoo are equivalent to those of
ISICS, the reengineering process determined major changes in the
software architecture and design, which have significant implications on
the use of its physics functionality in scientific applications. 

The reengineered version facilitates the exploitation of ISICS cross
section calculations for large scale productions requiring a variety of
settings, such as the tabulation of data libraries; it also enables the
exploitation of ISICS physics functionality within general-purpose Monte
Carlo systems for particle transport and PIXE (Particle Induced X-ray
Emission) analysis systems. More in general, the new form of ISICS is
suitable to applications where a versatile, platform-independent tool
for the calculation of inner shell ionization cross sections by proton
or ion impact is needed. As such, it complements the original ISICS
Windows version, which is oriented for small scale individual use.

The following sections outline the software design and the improvements
to the implementation available in this new development of ISICS; they also
document the use of the ISICSoo class and the test process it has been
subject to prior to its release in the CPC Library.

The reengineering of ISICS is based on its latest published version, ISICS 2008
\cite{isics2008}, complemented by an unpublished ISICS 2010 version
\cite{isics2010}, which includes a few small fixes to the 2008 version.
The changes in cross section calculations originally implemented in ISICSoo,
described in section \ref{ISICS_changes}, were successively implemented also in
the later ISICS 2011 version \cite{isics2011}.

\section{Software design}

The following sections provide an overview of relevant issues addressed by the
reengineering process and the resulting new features of the software design.

\subsection{The refactoring process}

Reengineering \cite{Demeyer2002} is a process of examination and
alteration of an existing software system to reconstitute it in a new
form. It exploits established techniques and patterns, that embody
knowledge and best practices adopted by the software community.

The reengineering of  ISICS addressed various issues, including:
\begin{itemize}
  \setlength{\itemsep}{1pt}
  \setlength{\parskip}{0pt}
  \setlength{\parsep}{0pt}
\item unbundling the monolithic ISICS system, so that some individual
  parts of it -- namely, its cross section calculations -- could be used
  independently; 
\item porting the system to other platforms than Microsoft Windows: this
  process required reworking the software architecture to clearly
  separate the platform-dependent code;
\item the absence of layering in the original procedural code hindered
  portability and adaptability: the reengineering analysis identified
  two layers - database and user interface - along with the proper
  physics domain;  
\item the shift from the procedural programming paradigm of ISICS to the
  object oriented paradigm. 
\end{itemize}

To a large extent, the final stage of the reengineering process
consisted of refactoring \cite{Fowler1999}, i.e. a process that changes
the structure of a software system without changing its observable
functionality. Refactoring methods specific to converting procedural
code to objects \cite{Fowler1999} were applied.

\subsection{Design features}

The adoption of the object oriented technology for the reengineering of
ISICS ensures a versatile, yet compact design. A single class, named
ISICSoo, is responsible for the physics functionality originally
provided by ISICS and for the interface with the client. 

In line with best practices of object oriented design
\cite{sutter101,cleancode,booch,wirfsbock}, ISICSoo responsibility is
focused on cross section calculation; other responsibilities in response
to user requirements (such as, for instance, error handling) are
expected to be provided by other entities (classes or packages) of the
user application or framework. 

Consistently with the object oriented paradigm, an ISICSoo object
encompasses functions associated with its responsibilities and the data
needed for fulfilling them.

The class constructor, invoked when an instance of the ISICSoo class is
created, is responsible for configuring the state of the instantiated
object with a set of default options, including the physics data to be
used in the calculations. 

The public class interface is minimal, being limited to operations
pertaining to interactions with the client, while the physics
functionality of the object is relegated to private member
functions. Public member functions set-up the calculation environment,
trigger the calculation of the requested cross sections and retrieve the
results of the computation. All data members are private.

The \lstinline{SetVerbosity} member function controls the textual output
from the ISICSoo object; it addresses different use cases, from batch
running in production mode to minute monitoring of the calculation.
Its integer argument defines the desired level of detail:
\lstinline{SetVerbosity(0)} suppresses all output,
\lstinline{SetVerbosity(1)} will print concise information, while
\lstinline{SetVerbosity(2)} will print exhaustive details. 

A Doxygen \cite{doxygen} documentation, illustrating the interface, data
members of the ISICSoo class and some examples, is included in the CPC
package.

\section{Client interface and computation configuration}

The class interface defines how a client can interact with an ISICSoo
object; it consists of functions for setting the parameters and the
input data to be used in the physics calculations.

In addition, the occurrence of anomalies in the execution of operations
of the ISICSoo class is signaled to the user by logging informative
error messages to the standard error stream \lstinline{std::cerr}, which
the user can redirect to a file, if desired.

\subsection{Configuration settings}

A number of parameters must be specified to set-up the computation
environment prior to performing cross section calculations: the
characteristics of the projectile (particle type and energy), the target
atom, the shell (or shells) for which the cross sections are to
be calculated, and the modeling approach to be activated (PWBA or
ECPSSR, the latter in turn articulated over a few different options).

The computation settings can be defined either through an external
configuration file, similarly to the original ISICS program, or through
public member function calls. The functionality for parsing
configuration files is delegated to private member
functions. Configurations settings defined through the ISICSoo public
interface can be saved to files for subsequent reuse. The class
interface provides convenient functionality to generate a large variety
of configurations, as may be needed for large scale productions (e.g. to
tabulate a cross section data library). The format of the configuration
files is compatible with the original ISICS 2008 version to facilitate
the reuse of existing settings in different environments.

ISICSoo provides the same ECPSSR options as the ISICS program: the
original formulation of the theory  \cite{ecpssr} (identified as ``plain
ECPSSR'' in the following), the United-Atom approximation
\cite{isics_ua}, the Hartree-Slater description of K-shell electrons
\cite{lapicki2005} and the treatment of K-shell ionization by
relativistic protons \cite{lapicki2008}. 

The calculation of cross sections for a given shell can be activated by
invoking the associated member function with the appropriate argument
(e.g.\lstinline{CalculateKShell(true)}). 

The projectile and the target can be specified either by passing the
atomic number $Z$ or the name of the element as an argument to the
member functions responsible for their definition.

In the reengineered version it is now possible to calculate the cross
sections at a single energy, at equidistant values within a given energy
range or by providing a list of predefined energies. 

Default values are attributed in the class constructor to all the
options selectable through the ISICSoo public interface. This feature,
which complies with good C++ programming practice, avoids the risk that
some variables -- namely data members -- may remain uninitialized. The
public class interface lets the user override the default values.

Default values provide a standard configuration of ISICSoo objects;
however, it is worthwhile to stress that optimal options for given
projectile and target parameters do not exist in absolute terms, but
depend on the use case. 

Moreover, differently from the limited spectrum of use of the stand-alone
ISICS program, several instances of the ISICSoo class may coexist in the
same application, each one characterized by different option settings:
examples of such use cases are a data library production, or the
investigation of different cross section models. 

Such flexibility is a relevant feature of the shift to the object
oriented paradigm and of the design of ISICSoo as a class, with respect
to the nature of procedural program of the original ISICS. The results
of ISICSoo validation process provide guidance for the selection of
options suitable to different experimental scenarios.

\subsection{Physics parameters }

The ISICSoo class uses various physics data in the calculation of cross
sections, such as masses of elements, atomic electron binding energies,
fluorescence and Coster-Kronig transition parameters.

The same data sets are supplied along with the source code as in the
previous version of ISICS \cite{isics2008}; the ISICSoo class can load
alternative data supplied by the user, provided the data files are
formatted in the same way as the original ones.

The data sets are expected to be found by default in the same directory
where ISICSoo resides; alternatively, the user can place them in a
different directory, whose path is specified through the
\lstinline{ISICS_DATA} environmental variable. An example of how to load
user-supplied data is given in the code listing  \ref{example_code}. 

\section{Features of ISICSoo implementation}
\label{c++}

\subsection{Generalities}
\label{sec_implemgen}

The ISICSoo class is implemented in C++; the choice of this language
allows the new version to profit from the benefits of the object
oriented paradigm, while the compliance of C++ syntax with C facilitated
refactoring the original C code. Indeed, large part of the original
ISICS C language code has been kept as is in the new class
implementation, changing the code  only where necessary to reflect the
new software design. 

The implementation of ISICSoo is compliant with C++ standard \cite{c++}
and uses the Standard Template Library, making the code highly portable
to different platforms.

In the process of migrating to C++, the precision of the calculation has
been changed from single precision (\lstinline{float}) to double
precision (\lstinline{double}), since double precision arithmetic is
nowadays considered a standard in mathematical calculations
\cite{NRC++}.  

The emphasis of this paper is on the calculation of ionization cross
sections; nevertheless ISICSoo also implements the calculation of X-ray
production cross sections for consistency with the legacy ISICS
code. According to the iterative-incremental software development
process adopted by ISICSoo, based on the Unified Process \cite{up}
framework, the responsibility for X-ray production is intended to be
delegated to a collaborating class (or package) in a future version.

\subsection{Changes in cross section calculations}
\label{ISICS_changes}

ISICSoo encompasses a few changes concerning the calculation of
ECPSSR cross sections with respect to the previous ISICS 2008 version,
which contribute to the robustness of the results and improve the
computational performance. 

\begin{figure}[ht]
\begin{center}
\includegraphics[width=0.78\textwidth]{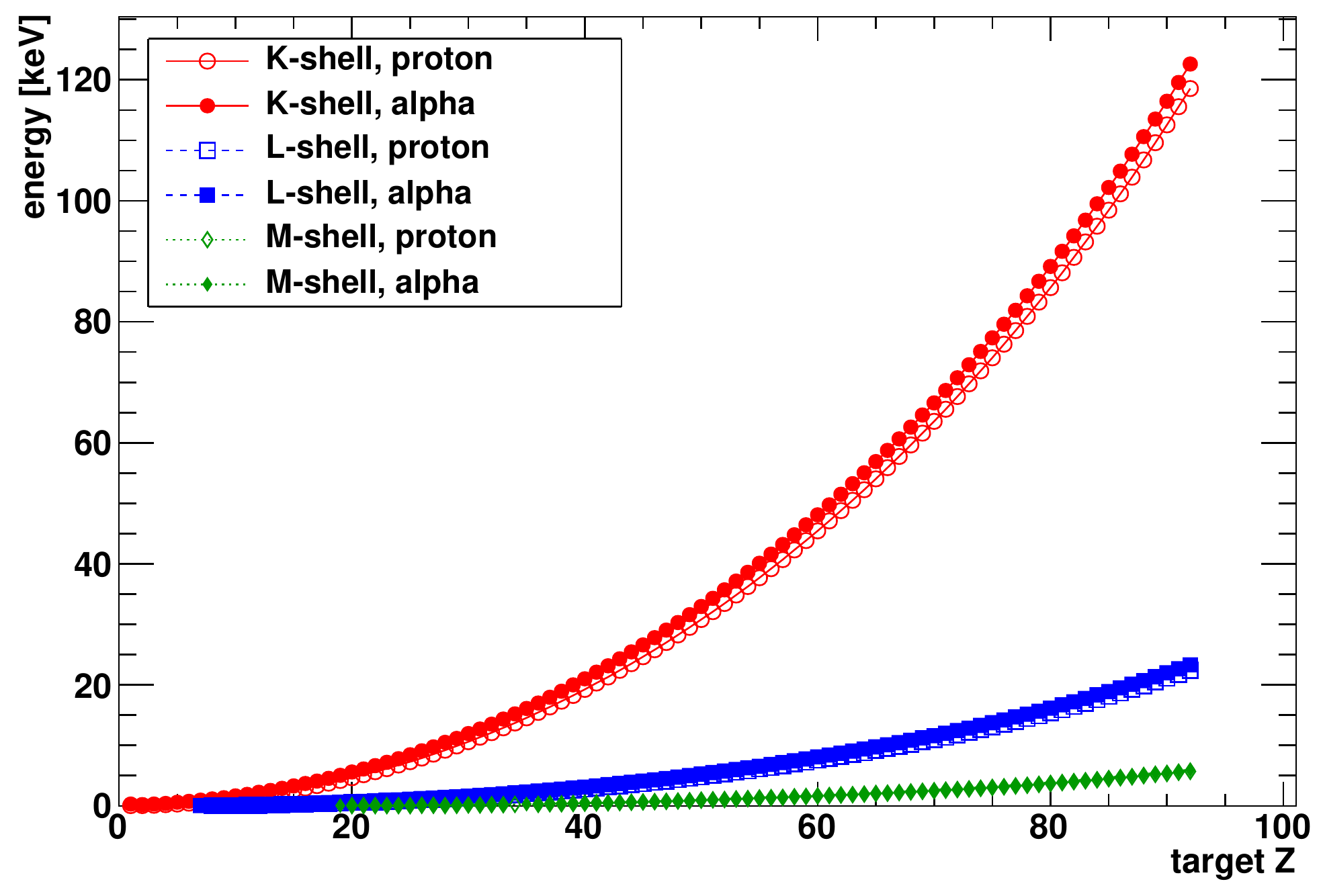}
\caption{Energy limit, below which negative values occur in the
expression of (\protect{\ref{eq:z_less_than_0}}), as a
function of the target element for different ionized shells and
projectiles as indicated in the legend.}
\label{fig:ecpssr_breakdown}
\end{center}
\end{figure}

Two of such changes had first been incorporated by the original author
of the ISICS program into ISICS 2010 version, which was downloaded from
the author's web page \cite{isics2010}. They prevent the generation of
unphysical results if the electron binding energy for a given shell
happens to equal zero in the data set used by ISICS, or the quantity
\begin{equation}
1-\frac{4}{M\zeta_S\theta_S}\left(\frac{\zeta_S}{\xi_S}\right)^2
\label{eq:z_less_than_0}
\end{equation}
which appears as the argument of a square root in the calculation of
$z_s$ (equation A.8 in \cite{isics}), is negative; in both cases the
cross section values are set to be zero. The upper energy limit, below
which the quantity in formula (\ref{eq:z_less_than_0}) becomes negative,
is illustrated in Figure \ref{fig:ecpssr_breakdown} as a function of the
target element for different ionized shells and protons and $\alpha$
particles as projectiles; one can observe that the occurrence of
unphysical negative values concerns relatively low energies with respect
to typical experimental applications of PIXE techniques, which represent
the main domain of interest for ISICS. 

The following other changes are specific to the ISICSoo implementation
and are not implemented in either 2008 or 2010 version of ISICS.

The test process adopted in the software development identified the
occurrence of unphysical negative limits passed to the integration
algorithm in two cases, which result in ECPSSR cross sections appearing
as \lstinline{nan} (not-a-number) in previous versions of ISICS. One of
such cases concerns a few light elements at low projectile energies,
when the $\zeta_S$ term of equation A.2 in \cite{isics} is negative. The
number of affected elements and energy range depends on the projectile;
for proton as projectile $\zeta_S$ is negative below $\sim 20$ keV for
oxygen and fluorine as target materials. 

The other case occurs when the binding energy for a given shell of the
combined target and projectile system (i.e. corresponding to an atom
whose atomic number is the sum of the atomic numbers of the two) happens
to be zero in the binding energy tabulations used by ISICS. This anomaly
has been observed, for instance, in M-shell ionization by proton impact
on bromine ($Z=35$), since the binding energy tabulations used by ISICS
report a value of zero for the M$_1$ binding energy of krypton
($Z=36$). The ISICSoo class returns zero as the value of the ECPSSR
cross section in both such cases.

In the special case of proton incident on bromine, the atomic binding
energy tabulation distributed with ISICSoo in the CPC Library has been
updated to include a positive value of the M$_1$ binding energy of
krypton, derived from \cite{carlson}. Special care should be taken to
ensure that user supplied binding energy tabulations provide positive
binding energy values not only for the target selected for calculation,
but also for the element corresponding to the composite target and
projectile system. 

Whenever, due to intrinsic deficiencies of the theory, numerical
artifacts or whatever other reasons, the cross section calculation
algorithm would result in unphysical values, the ISICSoo member
functions responsible for ECPSSR and PWBA cross section calculation
return zero value. Therefore, the objects interacting with ISICSoo
instances can be aware of the occurrence of unphysical conditions, while
dealing with such occurrences properly left to their responsibility. The
occurrence of unphysical cross section values is logged to the standard
error stream (\lstinline{std::cerr}). 

It is worthwhile to stress that, differently from ISICSoo, which is a 
stand-alone program, ISICSoo is a class and, as such, must have a single,
cohesive responsibility; other responsibilities in response to user
requirements (such as, for instance, error handling) are expected to be
provided by other entities -- classes or packages -- of the user
application or framework.

Other modifications with respect to the previous version of ISICS
concern the consistency of calculation of the available options of the
ECPSSR theory. These verifications prevent the inappropriate application
to the L-shell of the scaling function associated with the correction
for the relativistic Dirac-Hartree-Slater nature of K-shell, which
occurred in the previous version of ISICS.

Improved computational performance is achieved in ISICSoo by separating
the calculation of PWBA and ECPSSR cross sections in dedicated
functions. This modification prevents the unnecessary duplication of the
PWBA calculation, which occurs in the previous version of ISICS when
ECPSSR cross sections are calculated, and the redundant calculation of
ECPSSR cross sections, when only PWBA ones are desired.

As a further contribution to optimize the performance, special care has
been taken to ensure that the calculation of weights and abscissas for
Gaussian quadrature integration is performed just once, when the number
of polynomials is chosen through the client interface.

A new version of ISICS for Windows (ISICS 2011) \cite{isics2011} has
been released in the CPC Program Library after the submission of the
present paper to Computer Physics Communications. Its functionality is
equivalent to ISICSoo, as it incorporates the aforementioned changes
implemented in the ISICSoo class on top of the 2010 unpublished
version.

\section{Tests of the new version}
\label{validation} 

The reengineered version of the code has been subject to rigorous
testing, which includes the verification of the result of the
reengineering process and validation with respect to experimental
measurements. 

A simple application was developed for the purpose of testing, which
instantiates an object of ISICSoo type and uses it to calculate cross
sections corresponding to various modeling options in predefined
combinations of projectile and target parameters. 

\subsection{Verification}

The verification process involved the comparison of results calculated
by the ISICSoo class with the outcome of the original
ISICS. Consistently with the IEEE Standard for Software Verification and
Validation \cite{ieee1012}, the regression testing comprised in the
verification process was performed with respect to previous versions of
the code: the published ISICS 2008 version and the 2010 one, which can
be downloaded from the ISICS author's personal web site.

Apart from being the latest published version, ISICS 2008 was used to
produce a cross section data set distributed with Geant4
\cite{g4nim,g4tns} 9.4 version; therefore consistency of results is an
important verification for future data library productions based on
ISICSoo. ISICS 2010, although not released in the CPC Program Library,
was the original version in the reengineering process that led to
ISICSoo and verification of consistency with it was regularly monitored 
in the course of the refactoring process. 

The test concerned the plain ECPSSR K-shell ionization cross sections by
proton on elements with atomic number between 2 and 92, calculated for a
set of energies between 100 keV and 100 MeV. The calculations for
elements with missing or wrong binding energies data in the ISICS 2008
version (e.g. europium) were excluded; similarly, the calculation cases
affected by the modifications specific to ISICSoo described in section
\ref{ISICS_changes} were not considered in the regression tests with
respect to ISICS 2010. 

The relative differences between the cross section values calculated by
ISICSoo and ISICS (either 2008 or 2010) are below the machine precision
in $98.3 \pm 0.1 \%$ of test cases. The remaining $1.7 \pm 0.1 \%$ of
test cases exhibit relative differences up to approximately $10^{-3}$,
attributed to the different precision of calculation mentioned in
section \ref{sec_implemgen} and difference in number of significant
digits in the output between ISICS and ISICSoo. All such differences are
by far smaller than the experimental uncertainties by which inner shell
ionization cross sections are measured.   

As a result of the verification process, one can conclude that ISICSoo
cross section calculations are consistent with those of previous ISICS
versions; therefore, apart from the improvements described in
\ref{ISICS_changes}, the reengineering process has retained the same
correctness as the original ISICS physics output.  

It is worthwhile to remind the reader that the later ISICS 2011 version
is identical to ISICS 2010, apart from incorporating the changes
described in section \ref{ISICS_changes} originally implemented in
ISICSoo. Therefore, although reference \cite{isics2011} does not
explicitly document ISICS 2011 regression testing, on the basis of the
above verification process one can infer the equivalence of physics
results between ISICSoo and the later released ISICS 2011 version.

\subsection{Physics validation}
\label{performance_phys}

Software validation is an essential part of the software development
process and a key component in the assessment of the quality of a
software product. The validation of ISICSoo followed the guidelines of
the IEEE Standard for Software Verification and Validation
\cite{ieee1012}.

The validation process involved the comparison of cross sections by
proton impact calculated by ISICSoo with experimental measurements
\cite{paul_sacher,sokhi,orlic_exp}. This evaluation of compatibility of
the cross sections calculated by ISICSoo with experimental data provides
a quantitative appraisal of the accuracy achievable in applications of
the code.

The validation was limited to K and L-shell ionization, for which
extensive compilations of experimental measurements are available in the
literature; the scarcity of experimental data for M-shell ionization
prevented a similar analysis. The validation process adopted the same
method as documented in \cite{tns_pixe}. Due to the complexity of the
subject, only the main conclusions relevant to the assessment of
validity of ISICSoo are summarized here; the detailed results are
documented in a wider study \cite{tns_pcross} dedicated to the
evaluation of open source codes for the calculation of ionization cross
sections by proton impact. 

For each target element, cross sections were calculated at the same
energies as the experimental data for all the various ECPSSR options;
for the K-shell, cross sections were also calculated for the combined
Hartree-Slater, United-Atom and Relativistic-Projectile-Velocity
options. The compatibility of the calculated cross sections with
measurements was determined by means of the $\chi^2$ test; a 0.05
significance was set to reject the null hypothesis of compatibility
between calculation and experiment.   

For K-shell ionization, the null hypothesis is accepted with 0.05
significance in a fraction of test cases going from $67 \pm 6$\% to $77
\pm 5$\%, depending on the ECPSSR modeling option; the highest fraction
of compatible test cases corresponds to the ECPSSR with Hartree-Slater
correction configuration. For L-shell ionization, the best compatibility 
is achieved by cross sections calculated by ECPSSR with United-Atom
option; for this configuration the null hypothesis is accepted with 0.05
significance in $79 \pm 4$\% of the test cases over the three $L_1$,
$L_2$ and $L_3$ sub-shells. 

The validation process involved not only ISICSoo, but also two other 
freely available software systems for the calculation of ionization
cross sections based on the ECPSSR theory mentioned in section
\ref{intro}: ERCS08 and \v{S}mit's code. ISICSoo is found to achieve the
highest compatibility with experimental data over the whole set of K-
and L-shell test cases; the detailed results of the comparative
evaluation of the accuracy of the three codes are documented in
reference \cite{tns_pcross}. 

The validation process also evaluated possible systematic effects in the
cross sections due to the values of atomic electron binding energies
used in the calculation. This analysis examined the cross section
accuracy associated with the two options of binding energies distributed
with ISICS, the compilations by Bearden and Burr \cite{bearden} and by
Williams \cite{williams}, and with other compilations
\cite{eadl,carlson,toi1996,toi1978,sevier1979} used by the GUPIX PIXE
analysis code \cite{gupix1} and by general purpose Monte Carlo systems
for particle transport. Larger discrepancies with respect to
experimental data for K-shell are observed when using the atomic binding
energies collected in the Evaluated Atomic Data Library (EADL)
\cite{eadl}, while the cross section accuracy achieved with the other
compilations is statistically equivalent. The full set of results are
reported in detail in a wider scope paper \cite{tns_binding}, which
documents a survey of atomic binding energies compilations used by major
Monte Carlo transport systems.

The results of the ISICSoo validation process are also relevant to the
original ISICS, since the software verification process described in the
previous section established the equivalence of the cross section
calculated by ISICS and ISICSoo. It is worthwhile to note that the
validation of ISICSoo is more extensive than that of ISICS: for example,
although recent ISICS versions offer two options of atomic binding
energies, their relative effects on cross section accuracy were not
previously documented in association with ISICS. Users of both codes --
ISICSoo and original ISICS -- can profit from the outcome of the ISICSoo
validation process to optimize their choice of run configuration options
depending on their experimental requirements.

\section{Computational environment and performance}

The only prerequisite to use the ISICSoo class is a C++ compiler. 
The code has been tested on three platforms: Scientific Linux 5 with the
gcc 4.1.2 compiler, Mac OS X 10.6.5 with gcc 4.2.1 and Windows XP with
MS Visual C++ 2010 Express.

Using \lstinline{-O2} optimization with gcc version 4.1.2 took
approximately 3 times longer than compiling without optimization (about
6 seconds versus 2 seconds), but resulted in about 25\% shorter running
time. 

\begin{figure}[ht]
\begin{center}
\includegraphics[width=0.78\textwidth]{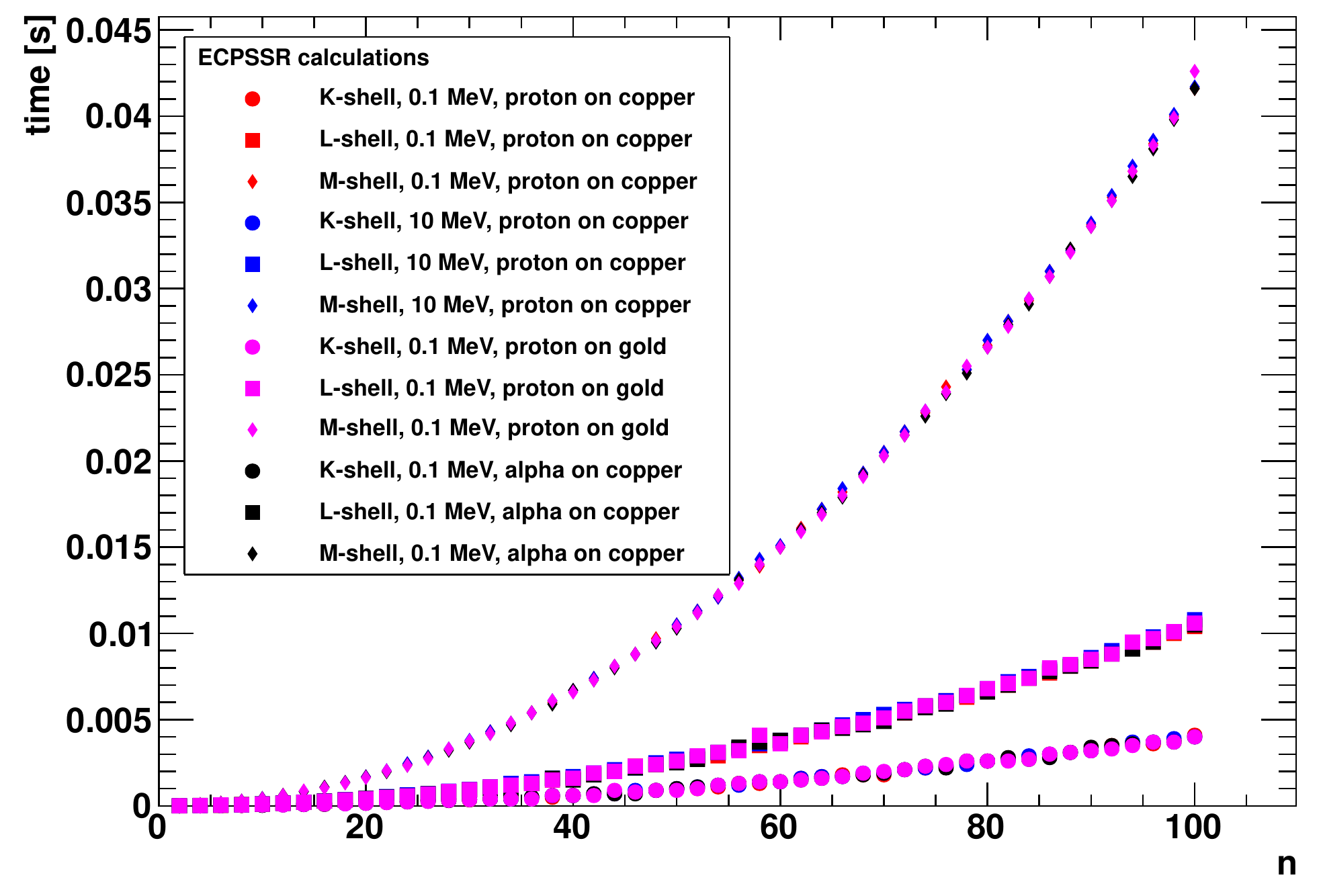}
\caption{Time to complete the calculation for various projectiles,
  targets, projectile energies and shells with respect to the
  Gauss-Legendre quadrature order. Circles are for K-shell, squares
  for L-shell and diamonds for M-shell calculations. Colors correspond
  to various projectiles, targets and energies, but the plot clearly
  shows that the calculations do not depend on them.} 
\label{fig:timing}
\end{center}
\end{figure}

Some timing tests have been performed on an AMD 2.4 GHz 2-core processor
machine, the results of which are shown in Figure
\ref{fig:timing}. There is no dependence on energy nor target or
projectile atomic number. The only changes in performance are due to
larger Gauss-Legendre quadrature order ($n$) used in integration and the
ionized shell. The calculation of L-shell  is about 3 times longer than
the calculation for K-shells, while for M-shell the calculation takes
approximately 10 times longer than for K-shell. 

\begin{figure}[ht]
\begin{center}
\includegraphics[width=0.78\textwidth]{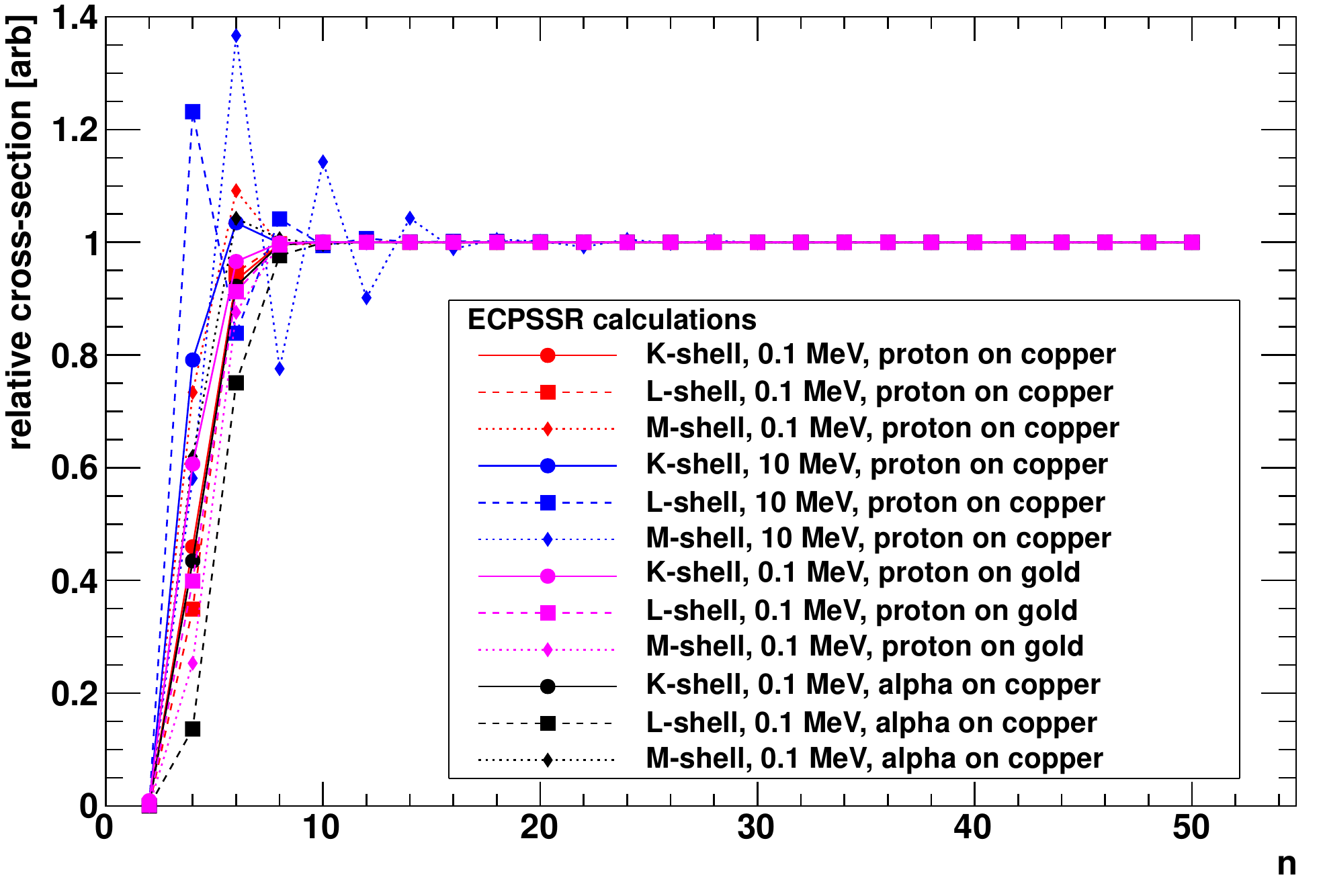}
\caption{The calculated cross section for ECPSSR calculated with
  respect to calculations at $n=100$ show that the results start
  converging at about $n=10$, depending on projectile energy.}
\label{fig:accuracy}
\end{center}
\end{figure}

Figure \ref{fig:accuracy} shows how the calculated cross-section
converge with increasing Gauss-Legendre quadrature order. The default
quadrature order ($n=50$) is not included in the plot, as the results
start converging already at about $n=20$. Hence, if timing performance
is of utmost priority, using $n=20$ or even $n=30$ will speed the
calculation, while the results will still be sufficiently precise. The
default value ($n=50$) is set in the class constructor; it can be
modified through the public interface of ISICSoo. As shown in figure
\ref{fig:accuracy}, the default value is conservative; if optimization
is needed by computationally intensive use cases as a trade-off between
accuracy and speed of execution, such a process should be done by the
interested user, since it may be affected by the computational
environment. However, it should be noted that this parameter affects the
precision of the cross section calculation, while the accuracy of cross
sections is determined by a variety of reasons, most of which depend on
the intrinsic physics capabilities of the theory, rather than on the
mathematical precision of the calculation.

Using various options of ECPSSR calculations (e.g.\ Hartree-Slater
correction) does not influence the running time significantly. 

\section{Application example}\label{example}

A simple example of how to use the ISICSoo class is illustrated in 
Listing \ref{example_code}. It consists of a main function, where an
instance \lstinline{isics} of ISICSoo is created (line \ref{p:inst}) and 
used to calculate the K-shell ECPSSR ionization cross section of 100 keV 
proton on gold. 

\begin{lstlisting}[caption={Example of a program using ISICS class.},label={example_code},escapeinside={@}{@}] 
#include <iostream>
@\label{p:include}@#include <ISICSoo.hh>
using namespace std;

int main(){
  @\label{p:inst}@ISICSoo* isics=new ISICSoo(); 

  @\label{p:data}@isics->LoadData("energy","energy_GW.dat");
  @\label{p:data2}@isics->LoadData("mass","my_mass.dat");

  @\label{p:opt1}@isics->SetN(50);
  @\label{p:opt2}@isics->SetProjectileZ(1);//proton
  @\label{p:opt3}@isics->SetTarget("Au");  //gold
  @\label{p:unit}@isics->SetEnergyUnit(1); //keV
  @\label{p:energy}@isics->SetEnergy(100);   //100 keV
  @\label{p:verbosity}@isics->SetVerbosity(1); 

  @\label{p:shell}@isics->CalculateKShell(true);
  @\label{p:hsr}@isics->SetHSRScaling(true);
  @\label{p:run}@isics->RunECPSSR();
  @\label{p:result}@double result=isics->GetShellECPSSR(0);

  @\label{p:print}@cout<<"K-shell ECPSSR cross-section for " 
      <<isics->GetProjectile()
      <<" with energy 100 keV on "
      <<isics->GetTarget()
      <<" is " 
      <<result<<" barn."<<endl; 
  return 0; 
}
\end{lstlisting}

The example shows how to replace default physics data with user selected
one in the cross section calculation: here Williams' atomic binding
energies (line \ref{p:data}), which are part of ISICS distribution, and
user supplied atomic masses (line \ref{p:data2}) are loaded. These data
files are expected to be in a directory identified through the
\lstinline{ISICS_DATA} environmental variable, or in the directory where
ISICSoo resides, if no environmental variable is specified.

Lines \ref{p:opt1}-\ref{p:verbosity} set some of the options for running
the calculation: the number of Legendre polynomials used in Gaussian
quadrature integration (line \ref{p:opt1}), projectile and target atomic
numbers (lines \ref{p:opt2} and \ref{p:opt3}), energy units for energy
input (0 for eV, 1 for keV and 2 for MeV) on line \ref{p:unit} and
energy of the projectile on line \ref{p:energy}. As described
previously, the \lstinline{SetVerbosity} function on line
\ref{p:verbosity} controls the textual output of the calculations. 
On line \ref{p:shell} the shell, for which calculations will be done, is
chosen; in this example this is the K-shell
(\mbox{\lstinline{CalculateKShell(true)}}). The user may choose to
calculate just one or several shells at the same time. 

The following line  \ref{p:hsr} selects the Hartree-Slater scaling
option for ECPSSR calculation.

Alternatively, the settings defined through ISICSoo public interface can 
be loaded from a configuration file by replacing lines
\ref{p:data}-\ref{p:hsr} with
\lstinline{isics->ReadConfig("test.cfg")}, where the function argument
corresponds to the file to be loaded. The configuration file
corresponding to the settings in this example is shown in the code
excerpt \ref{example_config}. 

The calculation of the ECPSSR model is initiated on line \ref{p:run}. 
The \lstinline{RunECPSSR()} member function will run the ECPSSR
calculations and populate the output data holders, one of which is then
retrieved on line \ref{p:result}, getting the ECPSSR cross section for
K-shell (index 0). 

\lstset{%
  morecomment=[l]{;},
  basicstyle=\footnotesize,%
  numbers=none,%
  keywordstyle=\color{black},%
  commentstyle=%
}
\begin{lstlisting}[caption={An example of a config file.},label={example_config}]
number of points for gauss legendre quadrature 50,
; this is comment
projectile atomic number 1, ;proton
target atomic number 79,    ;gold
projectile energy unit 1 (keV),
start energy 100 keV,
end energy 101 keV,
energy step 5 keV,
calculate K shell 1 (yes),
calculate L shell 0 (no),
calculate M shell 0 (no),
relativistic projectile velocity 0 (no),
united atom approximation 0 (no),
hsR scaling for K shell 1 (yes),
Run it (no) 0 ; I will call Run myself
\end{lstlisting}

Further examples are available in the \lstinline{examples} folder of the
package distributed in the CPC Program Library . 

\section{Conclusions}
\label{conclusions}

ISICS has been reengineered into a C++ class for easier portability
across different operating systems and to exploit its physics
functionality within other software systems, rather than as a
stand-alone program. 

Apart from the change in the software design, the new code includes 
some modifications to the software implementation, which improve the
robustness of the results and the computational performance.

A rigorous test process has verified the equivalence of the reengineered
version with respect to the previous one (apart from the above mentioned 
improvements) and has evaluated the compatibility of ISICSoo
calculations with an extensive collection of experimental data. The
implementation improvements originally developed in ISICSoo have been
introduced in the ISICS 2011 version to ensure equivalent functionality
of the stand-alone code on Windows platforms with respect to ISICSoo. The
achievable accuracy has also been favorably compared to the
results of other similar codes for the computation of ECPSSR cross
sections.  

The reengineered version of ISICS has been successfully exploited to
produce the PIXE data library \cite{pixe_datalib} distributed by RSICC,
which is used for PIXE simulation \cite{tns_pixe} in the Geant4 toolkit 
\cite{g4nim,g4tns}. 

Concepts and methods exploited in reengineering ISICS into ISICSoo could
be also useful to the evolution of other stand-alone programs available
in the CPC Library into components usable in larger scale software
systems.

\bibliographystyle{model1a-num-names} 

\end{document}